\documentclass[journal,oneside]{IEEEtran}
\usepackage{amsmath,amsfonts}
\usepackage{graphicx}
\usepackage{booktabs,array}
\usepackage{cite,url}
\usepackage{algpseudocode}
\usepackage{algorithm}
\usepackage{multicol}

\begin{document}

\title{Low-Complexity System and Algorithm for an Emergency Ventilator Sensor and Alarm}
\author{
    \IEEEauthorblockN{Ryan~M.~Corey\IEEEauthorrefmark{1},
    Evan~M.~Widloski\IEEEauthorrefmark{1},
    David~Null\IEEEauthorrefmark{1},
    Brian~Ricconi\IEEEauthorrefmark{2},
    Mark~Johnson\IEEEauthorrefmark{3},
    Karen~White\IEEEauthorrefmark{3},
    Jennifer~R.~Amos\IEEEauthorrefmark{4},
    Alex~Pagano\IEEEauthorrefmark{5},
    Michael~Oelze\IEEEauthorrefmark{1},
    Rachel~Switzky\IEEEauthorrefmark{6},
    Matthew~B.~Wheeler\IEEEauthorrefmark{4}\IEEEauthorrefmark{7},
    Eliot~Bethke\IEEEauthorrefmark{8},
    Clifford~Shipley\IEEEauthorrefmark{9},
    and~Andrew~C.~Singer\IEEEauthorrefmark{1}}
    
    \IEEEauthorblockA{\IEEEauthorrefmark{1}Electrical and Computer Engineering, University of Illinois at Urbana-Champaign},
    \IEEEauthorblockA{\IEEEauthorrefmark{3}Carle Foundation Hospital},
    \IEEEauthorblockA{\IEEEauthorrefmark{2}Creative Thermal Solutions, Inc.},
    \IEEEauthorblockA{\IEEEauthorrefmark{4}Bioengineering, University of Illinois at Urbana-Champaign},
    \IEEEauthorblockA{\IEEEauthorrefmark{5}Mechanical Science and Engineering, University of Illinois at Urbana-Champaign},
    \IEEEauthorblockA{\IEEEauthorrefmark{6}Siebel Center for Design, University of Illinois at Urbana-Champaign},
    \IEEEauthorblockA{\IEEEauthorrefmark{7}Animal Sciences, University of Illinois at Urbana-Champaign},
    \IEEEauthorblockA{\IEEEauthorrefmark{8}Carle Illinois College of Medicine},
    \IEEEauthorblockA{\IEEEauthorrefmark{9}Veterinary Clinical Medicine, University of Illinois at Urbana-Champaign}
    \thanks{This research was supported by the Grainger College of Engineering and the Siebel Center for Design at the University of Illinois at Urbana-Champaign, by Carle Health, and by an appointment to the Intelligence Community Postdoctoral Research Fellowship Program at the University of Illinois at Urbana-Champaign, administered by Oak Ridge Institute for Science and Education through an interagency agreement between the U.S. Department of Energy and the Office of the Director of National Intelligence.}%
}

\markboth{Corey \MakeLowercase{\textit{et al.}}: Low-Complexity System and Algorithm for an Emergency Ventilator Sensor and Alarm}{}

\maketitle

\begin{abstract}
In response to the shortage of ventilators caused by the COVID-19
pandemic, many organizations have designed low-cost emergency ventilators.
Many of these devices are pressure-cycled pneumatic ventilators, which are easy
to produce but often do not include the sensing or alarm features found
on commercial ventilators. This work reports a low-cost, easy-to-produce
electronic sensor and alarm system for pressure-cycled ventilators
that estimates clinically useful metrics such as pressure and respiratory
rate and sounds an alarm when the ventilator malfunctions. A low-complexity
signal processing algorithm uses a pair of nonlinear recursive envelope
trackers to monitor the signal from an electronic pressure sensor
connected to the patient airway. The algorithm, inspired by those
used in hearing aids, requires little memory and performs only a few
calculations on each sample so that it can run on nearly any microcontroller. 
\end{abstract}

\section{Introduction}

\IEEEPARstart{T}{he} COVID-19 crisis may cause shortages of ventilators used to treat patients with severe respiratory symptoms. 
COVID-19 patients can experience acute respiratory distress syndrome (ARDS), which causes extreme difficulty breathing due to fluid leaking into the lungs \cite{alhazzani2020surviving,zhang2020evolving, grasselli2020baseline}. Mechanical ventilation can help to treat these patients by providing oxygen while the underlying disease runs its course \cite{alhazzani2020surviving,matthay2020treatment}. Appropriate oxygen delivery is a mainstay of critical care and in COVID-19 can prevent death from ARDS and hypoxemia. 

Because the growing number of COVID-19 cases may exceed the number of available ventilators, dozens
of companies, university research teams, and other organizations have developed
emergency ventilators under special authorizations from regulators \cite{fda2020enforcement,mrha2020rapid}. Pressure-cycled pneumatic ventilators, like the Illinois
RapidVent developed by the authors' institutions \cite{rapidvent}, are especially attractive for this emergency because they can be rapidly and inexpensively manufactured. They are powered by pressurized gas and controlled by a mechanical modulator, so they require no electronic components to operate. 
However, they lack the sensors found in more-expensive commercial ventilators that provide closed-loop control, monitoring, and alarm capabilities. 
Clinicians rely on these electronic systems to adjust ventilator settings and to alert them to ventilator malfunctions or patient activity that require their attention.
Without sensing and alarm features, clinicians must constantly monitor
each patient and cannot be sure that ventilator settings are correct.

This work describes an electronic sensor and alarm system for pressure-cycled
emergency ventilators. Like the pneumatic ventilators it is designed
to complement, this device must be of low cost and must be easy to
produce from readily available components. The most important function of the device is to sound
an alarm when the breathing cycle is abnormal. Because pressure-cycled
ventilators use pressure levels to switch between inhalation and exhalation
modes, they produce distinctive pressure waveforms \cite{hess2005ventilator,lian2009understanding}. The sensor
and alarm system can analyze this pressure signal to determine whether
the ventilator is cycling normally. The same pressure signal can be used to
detect sudden pressure loss due to disconnection and pressure spikes
due to mechanical failure and to estimate clinically useful parameters
such as the peak inspiratory pressure (PIP), positive end-expiratory
pressure (PEEP), and respiratory rate (RR). 

\begin{figure}
    \centering\includegraphics[width=8cm]{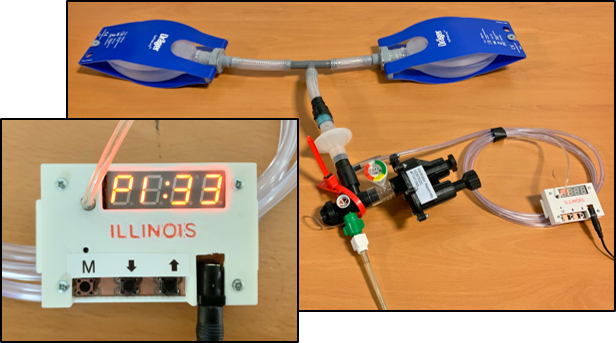}
    \caption{\label{fig:photo}Prototype of the Illinois RapidAlarm sensor and
    alarm system attached to a pressure-cycled ventilator and artificial lung.}
\end{figure}

These sensing and alarm functions reproduce most of the functionality of similar commercial monitoring products. 
A full-featured monitoring system would also measure tidal volume, or the amount of air delivered with each breath, and the oxygen concentration in the air, and it would trigger alarms based on these metrics \cite{mrha2020rapid}.
However, tidal volume and oxygen concentration cannot be inferred from the pressure waveform alone and would require more complex equipment.

The device reported here, known as the Illinois RapidAlarm and shown
in Fig. \ref{fig:photo}, monitors the pressure-cycled ventilator
using an electronic pressure sensor that connects to the patient airway
using a standard respiratory tubing adapter. A microcontroller analyzes
the pressure signal using low-complexity signal processing algorithms
inspired by audio processing methods used in hearing aids. Because
the proposed algorithm does not store past samples of the signal in
memory and performs only a few calculations on each sample, it can
run on nearly any microcontroller. The hardware design files and software
code for the Illinois RapidAlarm are available online under open-source
licenses\footnote{https://rapidalarm.github.io}. This work describes
the design of the system, with particular attention to the signal
processing algorithm used to estimate breathing metrics and detect
malfunctions. The algorithm is validated using animal data and a hardware prototype is demonstrated using an artificial lung.

\section{Pressure-Cycled Ventilation}

\begin{figure}
    \centering\includegraphics{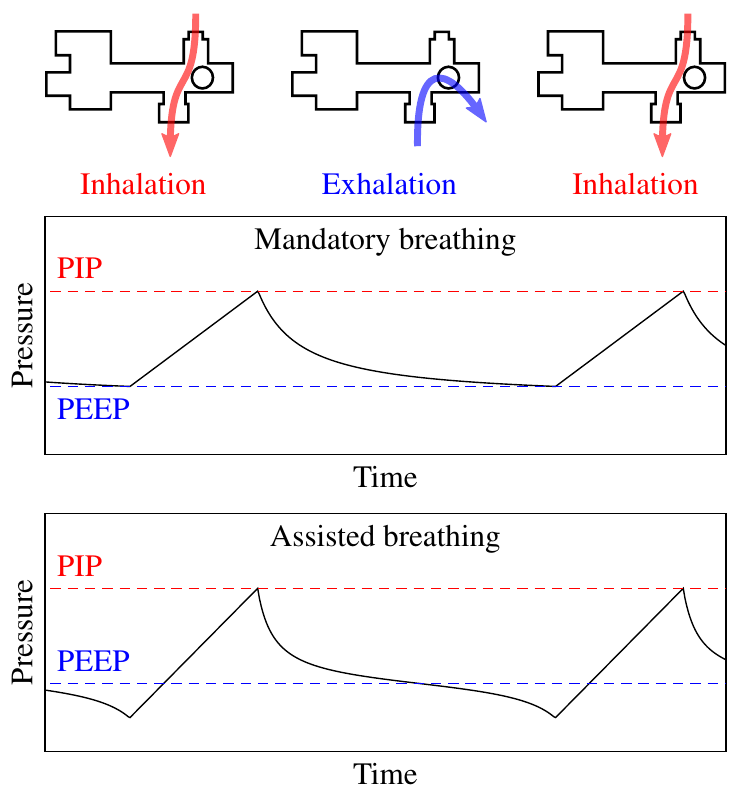}
    \caption{\label{fig:pressure_cycled}A pressure-cycled ventilator uses positive pressure to deliver gas to the patient airway. During normal operation, it produces a distinctive pressure waveform.}
\end{figure}

Pressure-cycled pneumatic ventilators, which are powered by pressurized gas, are useful
in the present emergency because they have low cost, are easy to
manufacture, and require no electronic components for basic operation \cite{lher2014vents}.
They provide pressurized gas to the patient airway and cycle between inhalation and exhalation modes using a pressure-switching mechanism controlled by pneumatic logic, as shown in Fig. \ref{fig:pressure_cycled}. 
During inhalation, high-pressure gas flows from the ventilator to the patient's lungs. As the lungs inflate, the pressure in the airway increases until it reaches the peak inspiratory pressure (PIP), a maximum pressure threshold that can be adjusted by the user.

Once PIP is reached, the modulator opens a path to the atmosphere that allows air from the lungs to exit the
ventilator. 
During exhalation, the pressure in the airway drops steadily, but does not fall to atmospheric pressure. Instead, once it drops below the positive end-expiratory pressure (PEEP), a spring closes the path to the atmosphere to initiate the next inhalation. During assisted-breathing mode, also known as pressure-support mode, the patient initiates a breath by inhaling to pull the pressure below the PEEP threshold. The clinician can control flow by adjusting a PIP dial and a rate dial, which determines expiratory time.

In pneumatic ventilators, the PEEP threshold is a fixed fraction of PIP determined by the mechanical design of the device.
Because COVID-19 patients can require PEEP levels in the range 10-15 cm H\textsubscript{2}O and PIP levels in the range 30-40 cm H\textsubscript{2}O \cite{grasselli2020baseline}, some COVID-19 emergency ventilators are designed with smaller PIP-to-PEEP ratios than commercial ventilators.
For example, the Illinois RapidVent has a measured PIP-to-PEEP ratio of about 2.4.

Pressure-cycled ventilators produce characteristic pressure waveforms,
as shown in Fig. \ref{fig:pressure_cycled}. During mandatory breathing,
the airway pressure rises from PEEP to PIP during inspiration, then
drops from PIP to PEEP during exhalation. During assisted breathing,
the basic shape is the same, but the pressure may fall below PEEP when the patient initiates inhalation.
The pressure signal can be used to estimate the PIP, PEEP (or minimum
pressure for assisted breathing), and RR.

\begin{figure}
    \centering\includegraphics{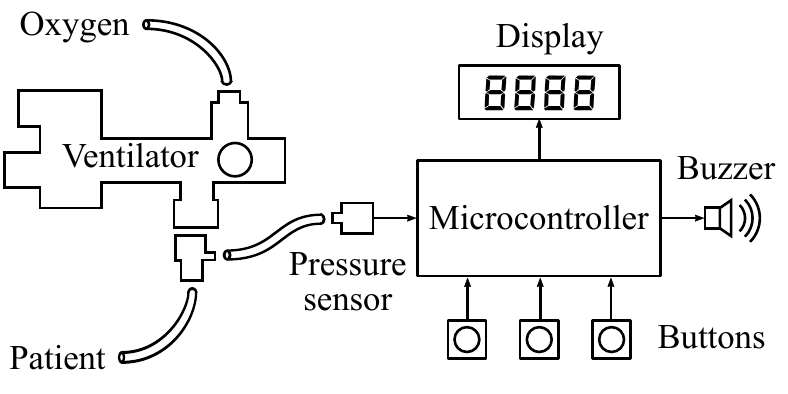}
    \caption{\label{fig:system}An electronic pressure sensor monitors gas pressure
    in the patient airway. A low-power microcontroller processes the pressure
    signal to generate measurements and alarms.}
\end{figure}

Because pressure-cycled ventilators produce well-defined pressure
waveforms during normal operation, the pressure signal can also be
used to detect malfunctions. If the gas circuit becomes obstructed
or disconnected, the modulator will stop cycling between inhalation
and exhalation modes, causing the pressure signal to remain constant.
The proposed alarm system uses low-complexity signal processing algorithms
to detect this constant-pressure condition.

\section{Sensing System}

To ensure that it is broadly useful in as many settings as possible,
the sensor and alarm system is not designed to be integrated with
any particular model of ventilator. Rather, it is a standalone component
that can be attached to any pressure-cycled ventilator. Because it
is intended to address the present emergency shortage, the design
prioritizes cost and ease of production. The system uses low-cost,
widely available parts, can be assembled on a two-layer printed circuit
board using either through-hole or surface-mount components, and runs
on a standard 5 volt power supply.

The sensor and alarm system is shown in Fig. \ref{fig:system}. The
device connects to the patient airway using standard respiratory tubing
adapters attached on the patient side of the ventilator. The electronic
system consists of a microcontroller, a display module, push buttons,
a buzzer, and a pressure sensor. Our implementation uses the 8-bit microcontroller ATmega328, which was selected for its ease
of use and wide availability. It is driven by an internal 8MHz clock and does not require an external oscillator. Because the computational requirements
of the proposed algorithm are low, as explained in Sec. \ref{sec:implementation},
nearly any microcontroller with an analog-to-digital converter and
several digital inputs and outputs should
be suitable for the sensor and alarm module. The open-source firmware
provides a hardware-agnostic C implementation of the monitoring algorithm that can be ported to other systems.

The user interface consists of three buttons and a four-character
seven-segment display. In display mode, the display cycles through
the three metrics (PIP, PEEP, and RR) every few seconds. The buttons are used to enable
and disable the alarm and to adjust user-configurable alarm settings,
which are described in Sec. \ref{sec:alarms} and summarized in Table \ref{tab:alarms}.
The alarm itself is a 4 kHz piezoelectric buzzer.

The key component in the system is the pressure sensor. The sensor interfaces with
the patient airway via a tube and converts pressure levels into electrical
signals, which are transmitted to the analog-to-digital converter
on the microcontroller. To capture the range of pressure levels produced
by pressure-cycled ventilators, the sensor should have a range of
at least 0 to 50 cm H\textsubscript{2}O. In our implementation, we used the NXP MPXV5010
piezoresistive pressure sensor, which has a pressure range of about
0 to 100 cm H\textsubscript{2}O and provides output voltages from about 0 to 5 volts.

\section{Pressure Tracking}

\begin{figure}
    \centering
    \includegraphics{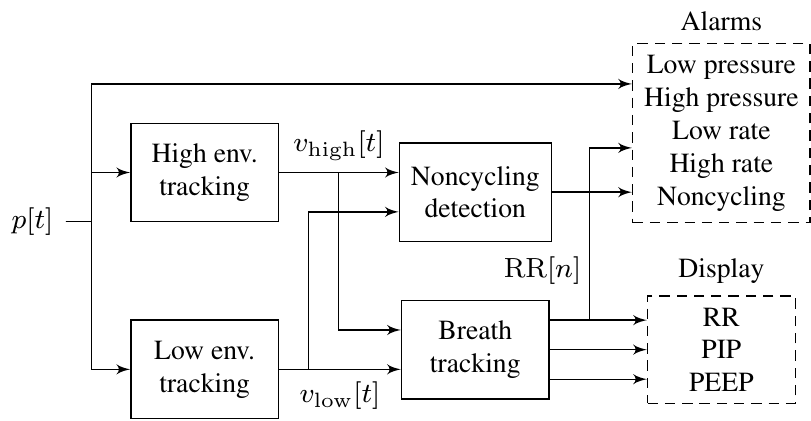}
    \caption{Alarm conditions and clinical metrics are derived from the measured pressure signal.}
    \label{fig:signal_processing}
\end{figure}

\begin{figure}
\centering\includegraphics{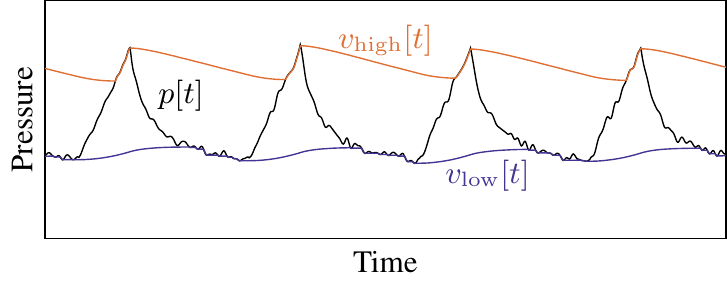}
\caption{\label{fig:envelopes}A pair of recursive peak detectors track the
envelope of the pressure signal without storing past samples in memory.}
\end{figure}

The behavior of pressure-cycled ventilators is well characterized
by the pressure signal measured at the patient airway. During normal
operation, the pressure cycles between PIP and PEEP once per breath,
as shown in Fig. \ref{fig:pressure_cycled}. In an ideal system, breaths
could be tracked by simply finding maxima and minima in this signal.
However, real signals do not always increase and decrease monotonically
like the waveform in that figure. The tracking algorithm must be robust
against small pressure variations and must have low computational
requirements so that it can run on inexpensive, low-power microcontrollers.

The proposed processing system, illustrated in Fig. \ref{fig:signal_processing}, uses a pair of nonlinear recursive filters to
track the envelope of the pressure signal. Recursive filters of the form $y[t]=ay[t-1]+bx[t]$ are widely used
in signal processing because of their computational efficiency: because they use feedback from the output to the input of the filter, they can perform many filtering tasks with less memory and fewer multiplications than feed-forward filters \cite{oppenheim1975digital}.
Envelope tracking uses a nonlinear version of this recursive filter: each performs a moving average, but gives more weight
to changes in one direction than another. The high-pressure envelope
increases quickly but decreases slowly, so it follows the top of the
pressure signal, while the low-pressure envelope decreases quickly
and increases slowly, following the bottom of the signal, as illustrated in Fig. \ref{fig:envelopes}.
Envelope detection is widely used as part of dynamic range compression in music production and
in digital hearing aids, which also have severe computational constraints \cite{giannoulis2012digital,kates2005principles}.

Let $p[t]$ be the discrete-time pressure signal from the sensor,
where $t$ is the sample index. The high-pressure envelope $v_{\mathrm{high}}[t]$
and the low-pressure envelope $v_{\mathrm{low}}[t]$ are given by
\begin{align}
v_{\mathrm{high}}[t] & \!=\!\begin{cases}
\alpha_{A}v_{\mathrm{high}}[t\!-\!1]+(1\!-\!\alpha_{A})p[t], & \!\text{if }p[t]\ge v_{\mathrm{high}}[t\!-\!1]\\
\alpha_{R}v_{\mathrm{high}}[t\!-\!1]+(1\!-\!\alpha_{R})p[t], & \!\text{if }p[t]<v_{\mathrm{high}}[t\!-\!1]
\end{cases} \label{eq:envelope_high}\\
v_{\mathrm{low}}[t] & \!=\!\begin{cases}
\alpha_{A}v_{\mathrm{low}}[t\!-\!1]+(1\!-\!\alpha_{A})p[t], & \!\text{if }p[t]\le v_{\mathrm{low}}[t\!-\!1]\\
\alpha_{R}v_{\mathrm{low}}[t\!-\!1]+(1\!-\!\alpha_{R})p[t], & \!\text{if }p[t]>v_{\mathrm{low}}[t\!-\!1],
\end{cases} \label{eq:envelope_low}
\end{align}
where $\alpha_{A}\in[0,1]$ and $\alpha_{R}\in[0,1]$ are called the
attack coefficient and release coefficient, respectively. These coefficients determine the relative importance of old and new samples in the moving average; they control how quickly the envelope tracker responds to changes in the pressure signal.

When the tracker is responding rapidly to a change in signal level
(an increase for the high-pressure envelope or a decrease for the
low-pressure envelope), it is said to be in attack mode. The attack
coefficient $\alpha_{A}$ is relatively small so that the tracker quickly forgets past estimates
and follows the new sample. For example, in our implementation at 100 samples/sec, $\alpha_A = 0.9$ so that the $1/e$ decay time of past samples in attack mode is around 10 ms. 

When the tracker is responding slowly
(to a decrease in pressure for the high-pressure envelope or an increase
for the low-pressure envelope), it is said to be in release mode.
The release coefficient $\alpha_{R}$ is closer to 1 so that the envelope decays more slowly,
allowing the algorithm to ignore small fluctuations
in pressure within a single breath cycle. The rate
of decay in release mode should be slow enough that the high and low envelopes stay
far apart during a normal breath cycle. If the tracker is too slow,
however, it could miss breaths when the pressure settings are adjusted
or, worse, might take too long to trigger an alarm when the ventilator
stops working. The choice of $\alpha_{R}$ is discussed further in
Sec. \ref{subsec:release_coeff}.

Notice that each recursive envelope tracker need only store one previous envelope value in memory. For comparison, a system using a rolling maximum/minimum filter approach at a sample rate of 100 samples per second would need to store about 200 past measurements for a window size of two seconds.

\section{Ventilation Monitoring}

\begin{figure}
\centering\includegraphics{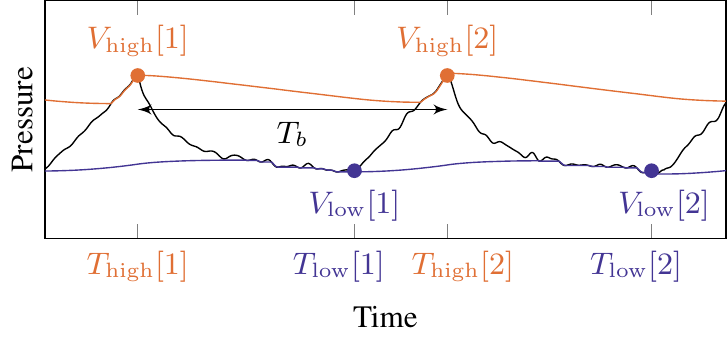}
\caption{\label{fig:breath_state}The inhalation/exhalation state of the ventilator
is inferred from attack events on the high-pressure and low-pressure
peak detectors.}
\end{figure}

The monitoring system estimates three metrics: PIP, PEEP (or
the minimum pressure of the breath cycle for assisted breathing),
and RR. All three of these metrics are tracked by detecting
inhalation and exhalation cycles from the pressure envelopes, as shown
in Fig. \ref{fig:breath_state}.

The two envelope trackers each store the most recent pressure sample
that triggered their attack mode, as shown in the figure. During each
inhalation cycle, there are several attack-mode samples in a row for
the high-pressure envelope. During exhalation, there are several attack-mode
samples in a row for the low-pressure envelope. The system tracks
breath cycles by looking for low-pressure attack events that follow
high-pressure attack events and vice versa. A low-pressure attack
event causes the system to switch from inhalation to exhalation mode,
and a high-pressure attack event causes it to switch from exhalation
to inhalation mode.

\subsection{PIP and PEEP}

When a mode switch occurs, the previous attack value is used to update
the corresponding PIP or PEEP estimate. That is, when a low-pressure
attack event occurs, the PIP display is updated with the most recent
high-pressure attack value. When a high-pressure attack event occurs,
the PEEP display is updated with the most recent low-pressure attack
value. Let $V_{\mathrm{high}}[n]$ and $V_{\mathrm{low}}[n]$ be the
peak values of the high- and low-pressure envelopes, respectively,
during breath cycle $n$, and let $T_{\mathrm{high}}[n]$ and $T_{\mathrm{low}}[n]$
be the sample indices at which they occur.

Both PIP and PEEP are recursively smoothed over time to remove small
fluctuations:

\begin{align}
\mathrm{PIP}[n] & =\alpha_{S}\mathrm{PIP}[n-1]+(1-\alpha_{S})V_{\mathrm{high}}[n]\\
\mathrm{PEEP}[n] & =\alpha_{S}\mathrm{PEEP}[n-1]+(1-\alpha_{S})V_{\mathrm{low}}[n],
\end{align}
where $\alpha_{S}$ is a smoothing coefficient between 0 and 1. This is a linear filter with an exponential impulse response; the contribution of sample $n_0$ to the moving average decays as $\alpha_S^{(n-n_0)}$. The closer $\alpha_S$ is to 0, the more quickly the display will respond to changes. We used $\alpha_{S}=0.5$ in our implementation.

\subsection{Respiratory rate}

\begin{figure}
\centering\includegraphics{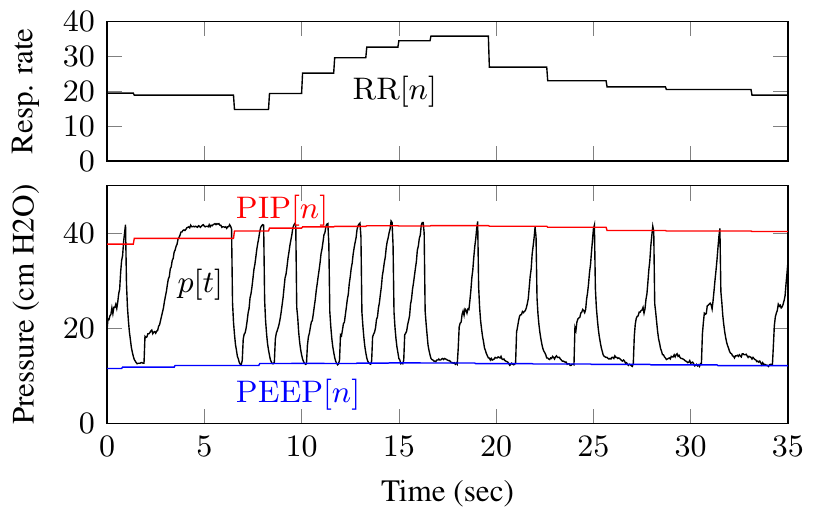}
\caption{\label{fig:experiment_measurements}Displayed PIP, PEEP, and respiratory rate values for experimental data from an artificial lung.}
\end{figure}

The system also keeps track of the time elapsed between these mode-switch
events. A complete breath cycle is measured between high-pressure
peaks. After smoothing, the average number of samples per breath is
\begin{equation}
    T_b[n] = \alpha_S T_b[n\!-\!1] + (1-\alpha_S)(T_\mathrm{high}[n]-T_\mathrm{high}[n\!-\!1]).
\end{equation}
Then the respiratory rate in breaths per minute is given by
\begin{equation}
\mathrm{RR}[n]=\frac{60 f_s}{T_b[n]},
\end{equation}
where $f_{s}$ is the pressure sensor sample rate in samples per second.
Note that although they are described mathematically as signals, in
practice $T_{\mathrm{high}}$ and $T_{\mathrm{low}}$ are implemented
as counters that reset on each breath cycle, as described in Sec.
\ref{sec:implementation}.

Figure \ref{fig:experiment_measurements} shows estimated PIP, PEEP,
and respiratory rate values superimposed on a pressure waveform measured
using the Illinois RapidVent connected to an artificial lung. Unlike the pressure envelopes, the displayed measurements are updated only once each breath cycle. Smoothing over multiple breaths prevents the values from changing too quickly, as observed in the respiratory rate plot.

\section{\label{sec:alarms}Alarm Conditions}

The monitoring device triggers alarms in several conditions that indicate the
ventilator is not working properly, as shown in Table \ref{tab:alarms}.
The alarm thresholds may vary between patients and between ventilator
devices and so they are configurable by the user. The table shows the range of values
that users of the Illinois RapidVent may select; these were chosen in consultation with local intensive-care experts.

\begin{table}
\caption{\label{tab:alarms}Alarm Conditions}
\setlength{\tabcolsep}{2pt}
\centering\begin{tabular}{@{}r c l@{}}
\toprule 
Alarm & Condition & Tunable range on RapidAlarm\\
\midrule 
High pressure & $p[t]>p_{\mathrm{max}}$ & $30\le p_{\mathrm{max}}\le90$ cm H\textsubscript{2}O\\
Low pressure & $p[t]<p_{\mathrm{min}}$ & $1\le p_{\mathrm{min}}\le20$ cm H\textsubscript{2}O\\
High RR & $\mathrm{RR}[n]>\mathrm{RR}_{\mathrm{max}}$ & $15\le\mathrm{RR}_{\mathrm{max}}\le60$ breath/min\\
Low RR & $\mathrm{RR}[n]<\mathrm{RR}_{\mathrm{min}}$ & $5\le\mathrm{RR}_{\mathrm{min}}\le15$ breath/min\\
Noncycling & %
\parbox{2cm}{
\begin{align*}
t-T_{\mathrm{high}}[n] & >T_{\mathrm{max}}\\
t-T_{\mathrm{low}}[n] & >T_{\mathrm{max}}\\
v_{\mathrm{high}}[t]/v_{\mathrm{low}}[t] & <r_{\mathrm{min}}\\
v_{\mathrm{high}}[t]\!-\!v_{\mathrm{low}}[t] & <d_{\mathrm{min}}
\end{align*}
} & $\,\,5\le \frac{T_{\mathrm{max}}}{f_s} \le30$ sec\tabularnewline
\bottomrule
\end{tabular}
\end{table}

\subsection{Pressure and respiratory rate}

The high- and low-pressure alarms trigger immediately if the sensor
detects a pressure outside the permitted range. In a pressure-cycled
ventilator, the pressure should never exceed the PIP value set by
the user. A pressure reading above the range of the PIP dial indicates
a mechanical failure. The low-pressure threshold $p_{\mathrm{min}}$
can be set close to zero, that is, atmospheric pressure, to detect
a disconnect in the breathing circuit. Note that because pressure-cycled
ventilators apply positive pressure even during exhalation, the airway
should never drop to atmospheric pressure unless the patient is attempting
to breathe spontaneously.

The high- and low-respiratory-rate alarms trigger if the average respiratory
rate falls outside the range specified by the user. A high respiratory
rate could indicate a low tidal volume, for example due to deteriorating
lung compliance, that requires a clinician's attention. The low-respiratory-rate
alarm has some overlap with the noncycling alarm, but it triggers
based on the average time between complete breath cycles, while the noncycling alarm
is triggered by the time elapsed since the last breath event.

\subsection{Noncycling conditions}

\begin{figure}
\centering\includegraphics{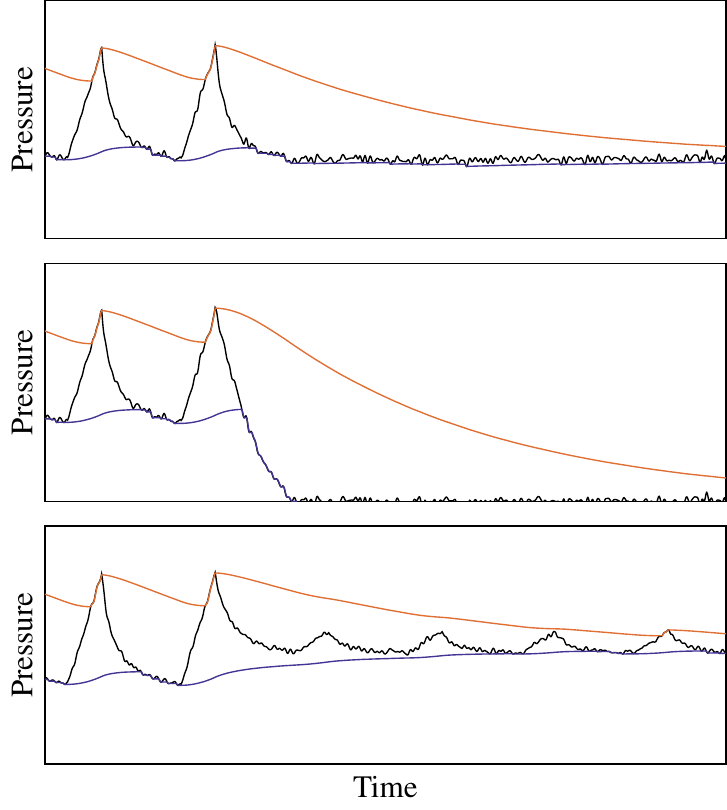}
\caption{\label{fig:noncycling}Different pressure signals that should trigger
a noncycling alarm.}
\end{figure}

The noncycling alarm condition is more complex than the first four.
It must detect when the breathing cycle has stopped, which can happen
in several ways, as illustrated in Fig. \ref{fig:noncycling}. Thus,
the alarm can be triggered by several conditions.

First, the alarm triggers if too much time has passed since the last
attack event of either envelope. For example, if the pressure drops
to PEEP and remains constant, as shown in the top panel of Fig. \ref{fig:noncycling},
there will be no attack events in the high-pressure envelope tracker,
so it will trigger the alarm. If, however, the pressure fluctuates
slightly over time, as shown in the bottom panel, the tracking algorithm will still detect frequent peaks.

To handle this case, the alarm will also trigger if the high-pressure
envelope and low-pressure envelope are too close together. In pressure-cycled
ventilators, the ratio between PIP and PEEP is a constant, here denoted
$r_{\mathrm{nom}}$, determined by the mechanical design of the device.
For the Illinois RapidVent, the nominal ratio is around 2.4. An alarm is triggered
if $v_{\mathrm{high}}[t]/v_{\mathrm{low}}[t]$ drops below $r_{\mathrm{min}}$,
a pressure-ratio threshold between 1 and $r_{\mathrm{nom}}$. The
alarm also triggers if the difference $v_{\mathrm{high}}[t] - v_{\mathrm{low}}[t]$
is too small. In our implementation, this minimum difference is fixed at 3 cm H\textsubscript{2}O.

\subsection{Parameter selection\label{subsec:release_coeff}}

\begin{figure}
\centering\includegraphics{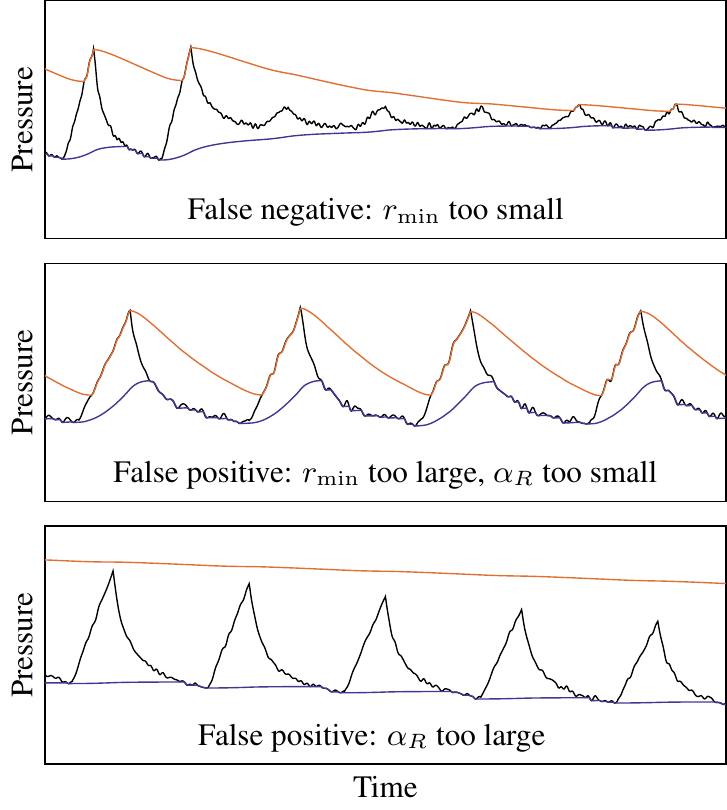}
\caption{\label{fig:errors}The parameters $\alpha_{R}$ and $r_{\mathrm{min}}$ must be carefully
chosen to avoid several types of detection errors.}

\end{figure}

\begin{figure*}
\centering\includegraphics{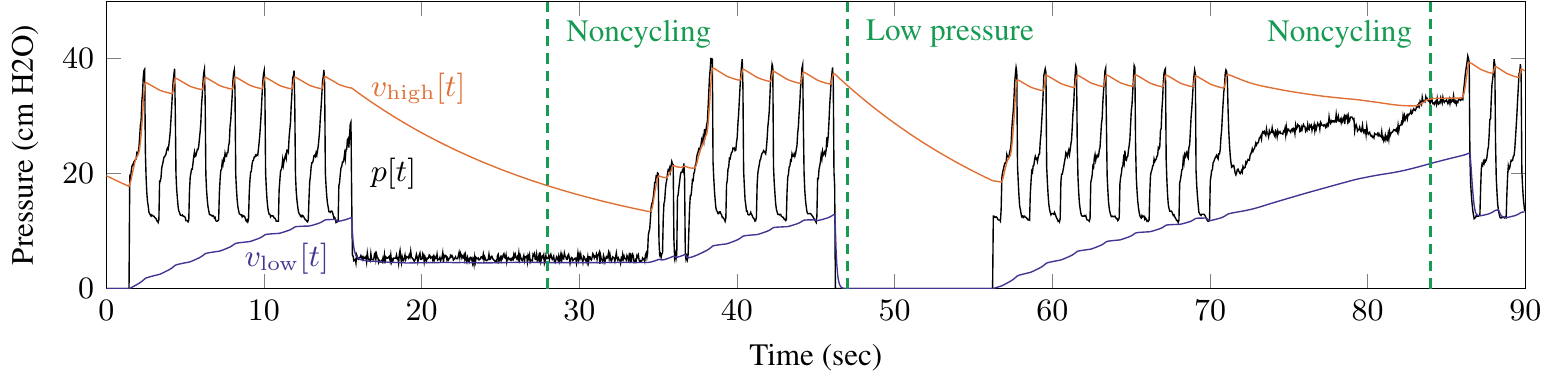}
\caption{\label{fig:experiment_alarms}Alarm-triggering conditions were simulated with an artificial lung. The dashed lines indicate alarm events.}
\end{figure*}

Because the pressure envelopes $v_{\mathrm{high}}[t]$ and $v_{\mathrm{low}}[t]$
naturally move toward each other during most of the breath cycle,
the release coefficient $\alpha_{R}$ and the high-to-low-pressure-ratio
threshold $r_{\mathrm{min}}$ jointly determine when the high-to-low-pressure-ratio
alarm will trigger. This alarm condition can detect several types
of ventilator malfunctions with low computational complexity. However,
it is also prone to false positives and false negatives and its parameters
must be carefully calibrated.

Figure \ref{fig:errors} shows three types of errors that could result
from poorly chosen values of $\alpha_{R}$ and $r_{\mathrm{min}}$.
First, if $r_{\mathrm{min}}$ is too small---that is, if it is close
to 1---the system could mistake noise in the pressure signal for
breath cycles. The alarm would then fail to trigger in the event of
an obstruction that prevents the ventilator from cycling but still
maintains a positive pressure. However, if $r_{\mathrm{min}}$ is
too large---that is, close to $r_{\mathrm{nom}}$---it could trigger
false alarms during normal breath cycles as the two envelopes fluctuate,
as shown in the middle of the figure.

The false-positive ratio alarm could be prevented by increasing the
value of $\alpha_{R}$ to make it closer to 1. A larger release coefficient
causes the envelopes to decay more slowly between breaths, so that
the ratio $v_{\mathrm{high}}[t]/v_{\mathrm{low}}[t]$ varies less over
the breath cycle. However, large values of $\alpha_{R}$ can impair
the ability of the system to adapt to changes in the pressure signal.
For example, if a clinician lowers the PIP setting using the dial
on the ventilator, as shown in the bottom panel of Fig. \ref{fig:errors},
the high-pressure envelope could miss the peaks of several breath
cycles. These missed attack-mode samples could falsely trigger the
time-since-last-peak alarm and would also cause errors in the PIP,
PEEP, and RR calculations.

These three cases illustrate the tradeoffs involved in the selection
of $\alpha_{R}$ and $r_{\mathrm{min}}$. To prevent the second error
type, the false alarm within a normal breath cycle, we can constrain
the relationship between the two parameters: as $r_{\mathrm{min}}$
increases, $\alpha_{R}$ must also increase. In our implementation,
we have set $\alpha_{R}$ so that, if the pressure were to fall suddenly
from PIP to PEEP and remain constant at PEEP, the noncycling alarm\textquoteright s
high-to-low-envelope-ratio condition ($v_{\mathrm{high}}[t]/v_{\mathrm{low}}[t]<r_{\mathrm{min}}$)
would be triggered at around the same time as its time-since-last-peak
condition ($t-T_{\mathrm{high}}[n]>T_{\mathrm{max}}$) for the default
alarm setting of $T_\mathrm{max}=15$ seconds. In this scenario, the high-pressure envelope decays
as 
\begin{equation}
v_{\mathrm{high}}[t]=\mathrm{PEEP}+\alpha_{R}^{t-T_{\mathrm{high}}}(\mathrm{PIP}-\mathrm{PEEP}).
\end{equation}
The alarm will therefore be triggered when 
\begin{equation}
v_{\mathrm{high}}[t]=r_{\mathrm{min}}\times\mathrm{PEEP},
\end{equation}
or 
\begin{equation}
r_{\mathrm{min}}=1+\alpha_{R}^{t-T_{\mathrm{high}}}\left(\frac{\mathrm{PIP}}{\mathrm{PEEP}}-1\right).
\end{equation}
Setting the elapsed time to the alarm time $T_{\mathrm{max}}$ and
solving for $\alpha_{R}$, we have 
\begin{equation}
\alpha_{R}=\left(\frac{r_{\mathrm{min}}-1}{r_{\mathrm{nom}}-1}\right)^{1/T_{\mathrm{max}}}.
\end{equation}

Having fixed $\alpha_{R}$ as a function of $r_{\mathrm{min}}$, we
must choose $r_{\mathrm{min}}$ to trade off between false positives
and false negatives. Choosing small values of $r_{\mathrm{min}}$
and $\alpha_{R}$ can lead to false negatives, as shown in the top
panel of Fig. \ref{fig:errors}. Choosing large values can lead to
false positives, as shown in the bottom panel. Because false negatives
are worse than false positives in a life-support device, $r_{\mathrm{min}}$
should be chosen to be comfortably larger than 1. In our implementation,
we use $r_{\mathrm{min}}=1.5$.

Figure \ref{fig:experiment_alarms} shows alarms triggered by experimental
data from the Illinois RapidVent connected to an artificial lung. The ventilator was obstructed and disconnected to create noncycling conditions. Notice that the low-pressure alarm triggers immediately when the breathing circuit is disconnected, while the noncycling alarm triggers after 15 seconds based on the user-specified alarm threshold.

\section{\label{sec:implementation}Algorithm Implementation}

\begin{figure}
\begin{algorithm}[H]
\caption{\label{alg:ventilator_monitoring}Ventilator monitoring algorithm}
\begin{algorithmic}
\small
\Loop
\State Read $p$ from pressure sensor
\State $T_\mathrm{peak} \gets T_\mathrm{peak}+1$
\State 
\If {$p \ge v_\mathrm{high}$}
	\State $v_\mathrm{high} \gets \alpha_A v_\mathrm{high} + (1-\alpha_A) p$
	\State $V_\mathrm{high} \gets p$
	\State $T_\mathrm{high} \gets 0$
	\If {breath state $=$ exhaling}
		\State breath state $\gets$ inhaling
		\State $\mathrm{PEEP} \gets \alpha_S \mathrm{PEEP} + (1-\alpha_S) V_\mathrm{low}$
	\EndIf
\Else
	\State $v_\mathrm{high} \gets \alpha_R v_\mathrm{high} + (1-\alpha_R) p$
	\State $T_\mathrm{high} \gets T_\mathrm{high} + 1$
\EndIf
\State 
\If {$p \le v_\mathrm{low}$}
	\State $v_\mathrm{low} \gets \alpha_A v_\mathrm{low} + (1-\alpha_A) p$
	\State $V_\mathrm{low} \gets p$
	\State $T_\mathrm{low} \gets 0$
	\If {breath state $=$ inhaling}
		\State breath state $\gets$ exhaling
		\State $\mathrm{PIP} \gets \alpha_S \mathrm{PIP} + (1-\alpha_S) V_\mathrm{high}$
		\State $\mathrm{RR} \gets \left(\alpha_S \mathrm{RR}^{-1} + (1-\alpha_S)\frac{T_\mathrm{peak} - T_\mathrm{high}}{60f_s}\right)^{-1}$
		\State $T_\mathrm{peak} \gets T_\mathrm{high}$
	\EndIf
\Else
	\State $v_\mathrm{low} \gets \alpha_R v_\mathrm{low} + (1-\alpha_R) p$
	\State $T_\mathrm{low} \gets T_\mathrm{low} + 1$
\EndIf
\State 
\State Check alarm conditions from Table \ref{tab:alarms}
\EndLoop
\end{algorithmic}
\end{algorithm}
\begin{itemize}
    \item $p$ -- current pressure sample
    \item $v_\mathrm{high}$ -- high-pressure envelope
    \item $v_\mathrm{low}$ -- low-pressure envelope
    \item $V_\mathrm{high}$ -- breath-cycle maximum
    \item $V_\mathrm{low}$ -- breath-cycle minimum
    \item $T_\mathrm{high}$ -- samples since most recent breath-cycle maximum
    \item $T_\mathrm{low}$ -- samples since most recent breath-cycle minimum
    \item $T_\mathrm{peak}$ -- samples since previous breath-cycle maximum
    \item $\mathrm{PIP}$, $\mathrm{PEEP}$, $\mathrm{RR}$ - smoothed PIP, PEEP, and RR output
    \item $\alpha_A$, $\alpha_R$ -- envelope attack and release coefficients
    \item $\alpha_S$ -- smoothing coefficient
    \item $f_s$ -- sample rate (samples per second)
\end{itemize}
\end{figure}

The monitoring algorithm can be implemented with low computational complexity and a small memory footprint.
The core loop of the algorithm is shown in Algorithm \ref{alg:ventilator_monitoring}.
The algorithm state comprises a binary inhalation/exhalation state variable; seven floating-point values, the envelopes $v_\mathrm{high}$ and $v_\mathrm{low}$, the breath-cycle peaks $V_\mathrm{high}$ and $V_\mathrm{low}$, and the estimated PIP, PEEP, and RR; and the three integer counters $T_\mathrm{peak}$, $T_\mathrm{high}$, and $T_\mathrm{low}$. 
The program must also store the user-configurable alarm thresholds in memory.

For every observed sample, the envelopes are updated according to \eqref{eq:envelope_high} and \eqref{eq:envelope_low} and the alarm conditions from Table \ref{tab:alarms} are checked.
Breaths are tracked using a state variable that toggles from inhalation to exhalation at the first low-pressure attack value in each breath cycle and from exhalation to inhalation at the first high-pressure attack value in each cycle. 
The time since the last attack for each envelope is tracked using counters $T_\mathrm{high}$ and $T_\mathrm{low}$. Thus, the time-based noncycling alarm conditions from Table \ref{tab:alarms} can be written $T_\mathrm{high}>T_\mathrm{max}$ and $T_\mathrm{low}>T_\mathrm{max}$.
An additional counter $T_\mathrm{peak}$ counts the samples since the previous high-pressure peak (the circles in Fig. \ref{fig:breath_state}), which allows the system to compute the time between breath cycles. 
The PIP, PEEP, and RR metrics are each updated once per breath cycle, regardless of the sample rate.

\begin{table}
\caption{\label{tab:utilization}Resource Utilization}
\centering\begin{tabular}{p{1.1cm} p{1.7cm} p{1.9cm} p{2.1cm}}
\toprule 
& Program storage (bytes) & Dynamic memory (bytes) & Execution time per sample (µs)\\
\midrule 
Interface & 6847 & 357 & 146\\
Algorithm & 4048 & 93 & 670\\
\midrule
Total & 10895 & 450 & 816\\
\bottomrule
\end{tabular}
\end{table}

\begin{table}
\caption{\label{tab:sample_rate}Sample Rate and Performance}
\centering\begin{tabular}{p{1.6cm} p{1.7cm} p{1.7cm} p{1.6cm}}
\toprule 
Sample rate (samples/sec) & RMS PIP error (cm H\textsubscript{2}O) & RMS RR error (breaths/min)\\
\midrule 
5 & 1.7 & 5.4 \\
10 & 0.5 & 0.4\\
20 & 0.6 & 0.2\\
50 & 0.3 & 0.1\\
100 & \multicolumn{2}{l}{Baseline} \\
\bottomrule
\end{tabular}
\end{table}

To assess the computational complexity of the system, the execution time of the algorithm was measured on the ATmega328 with a clock speed of 8 MHz. Table \ref{tab:utilization} shows storage, memory, and execution time of the monitoring algorithm and of the user interface logic that controls the buttons, buzzer, and display. The monitoring algorithm requires less than one millisecond per sample. The memory and storage requirements are dominated by the user interface logic.

Because the recursive envelope tracker performs a fixed number of calculations for each pressure sample, the overall computational complexity of the system depends on the sample rate.
To select an appropriate sample rate for the monitoring system, we must characterize the performance of the algorithm at different sample rates.
If the sample rate is too low, the sampled sequence might not capture the narrow peak of the pressure waveform, causing errors in the estimated PIP and RR values.

Table \ref{tab:sample_rate} shows the root-mean-square error in PIP and RR measurements when the algorithm is run at different sample rates. 
The pressure data is from an artificial lung cycling at a rate of more than 30 breaths per minute, which is faster than typical human respiratory rates. 
The error at lower sample rates is calculated relative to estimates performed at a baseline rate of 100 samples per second. 
Even for these fast breaths, the estimated PIP and RR values are accurate within the display resolution of 1 cm H\textsubscript{2}O and 1 breath/min for sample rates as low as 10 samples per second. 

These results suggest that the sensor and alarm module should use a sample rate of at least 10 samples per second. 
This rate is well within the capability of most modern microcontrollers, even low-cost 8-bit processors that must use many clock cycles to perform floating-point calculations.

\section{Animal Testing}

\begin{figure}
    \centering
    \includegraphics{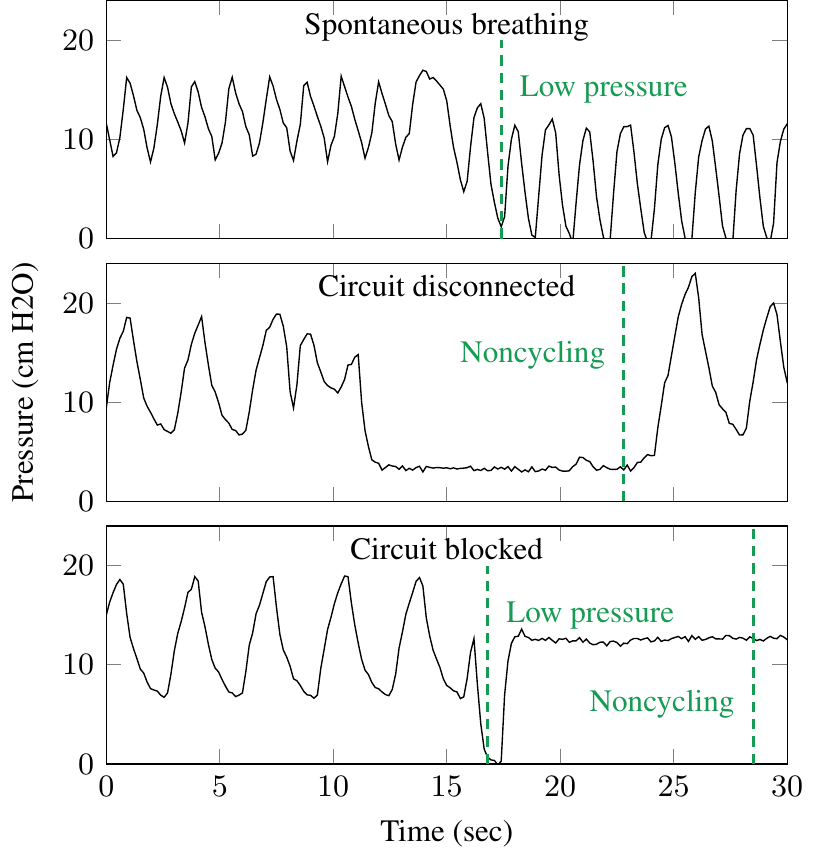}
    \caption{The monitoring algorithm was validated using data from a pressure-cycled ventilator on a sedated pig. Dashed lines indicate alarms triggered by the algorithm.}
    \label{fig:pig}
\end{figure}

The monitoring algorithm was validated using animal-testing data collected during the development of the Illinois RapidVent. The prototype ventilator was tested using sedated pigs, which have lungs that are similar in size to those of humans \cite{kuzmuk2011pigs,swindle2012swine}. The tests were conducted at the University of Illinois at Urbana-Champaign under protocol number 20071 approved by the Institutional Animal Care and Use Committee. Details are given in an upcoming paper about the Illinois RapidVent. The primary purpose of the tests was to evaluate the performance of the ventilator itself and the experiment did not include the prototype monitoring device. However, the pigs' airway pressure was measured continuously during the tests using a Rosemount 3051S differential pressure transducer sampled about seven times per second. The experiments included multiple combinations of PIP and rate dial settings, both mandatory and assisted breathing, and accidental and deliberate disconnections. Therefore, the pressure data captured during the animal tests are valuable for validating the proposed monitoring algorithm.

Figure \ref{fig:pig} shows excerpts from the pressure data and the alarms triggered by the monitoring algorithm. In the top panel, the pig inhaled strongly enough to pull pressure near zero (atmospheric level), triggering the low-pressure alarm. This alarm does not indicate ventilator failure, but alerts clinicians that the patient is breathing spontaneously. The middle panel shows a disconnection event: the pig rolled over, breaking the respiratory circuit. The pressure did not drop to atmospheric level but remained at a steady low level, triggering the noncycling alarm. In the bottom panel, the ventilator was deliberately blocked for several seconds in order to measure tidal volume, leading to a momentary drop and then steady high pressure level. This event triggers both the low-pressure and noncycling alarms. The envelope-tracking algorithm was found to work for all tested settings of the PIP and rate dials, although sudden changes in the dial settings can trigger false alarms and cause temporary inaccuracies in metric estimates.

\section{Conclusions}

The proposed sensor and alarm system can improve the functionality of pressure-cycled emergency ventilators. 
While it is not as robust as a full-featured commercial ventilator system, it provides critical monitoring features that are not available on purely mechanical ventilators. 
The recursive envelope-tracking algorithm allows the system to track breathing, estimate metrics, and detect malfunctions with only a few calculations per sample and a tiny memory footprint. 
Therefore, the system can be built quickly using nearly any low-cost microcontroller and a few other electronic components.

\end{document}